\documentclass[prx,twocolumn,showpacs,floatfix,amsmath,amssymb,superscriptaddress]{revtex4-1}
\usepackage{amsfonts}
\usepackage{stmaryrd}
\usepackage{bbm}
\usepackage{mathrsfs}
\usepackage{tipa}
\usepackage{amssymb}
\usepackage{txfonts}
\usepackage{graphicx}
\usepackage{dcolumn}
\usepackage{epstopdf}
\usepackage[colorlinks,linkcolor=blue,urlcolor=blue,citecolor=blue]{hyperref}
\usepackage{multirow}
\usepackage{subfigure}
\usepackage{url}

\begin{document}

\newcommand{\beq}{\begin{eqnarray}}
\newcommand{\eeq}{\end{eqnarray}}

\title{Magnetic excitations of the field induced states in BaCo$_{2}$(AsO$_{4}$)$_{2}$ probed by time-domain terahertz spectroscopy}

\author{L. Y. Shi}
\affiliation{International Center for Quantum Materials, School of Physics, Peking University, Beijing 100871, China}

\author{X. M. Wang}
\affiliation{International Center for Quantum Materials, School of Physics, Peking University, Beijing 100871, China}

\author{R. D. Zhong}
\thanks{Current address: Shanghai Jiao Tong University,
Shanghai 200240, China}
\affiliation{Department of Chemistry, Princeton University, Princeton, NJ 08544, USA.}

\author{Z. X. Wang}
\affiliation{International Center for Quantum Materials, School of Physics, Peking University, Beijing 100871, China}

\author{T. C. Hu}
\affiliation{International Center for Quantum Materials, School of Physics, Peking University, Beijing 100871, China}

\author{S. J. Zhang}
\affiliation{International Center for Quantum Materials, School of Physics, Peking University, Beijing 100871, China}

\author{Q. M. Liu}
\affiliation{International Center for Quantum Materials, School of Physics, Peking University, Beijing 100871, China}

\author{T. Dong}
\affiliation{International Center for Quantum Materials, School of Physics, Peking University, Beijing 100871, China}

\author{F. Wang}
\email{wangfa@pku.edu.cn}
\affiliation{International Center for Quantum Materials, School of Physics, Peking University, Beijing 100871, China}

\author{N. L. Wang}
\email{nlwang@pku.edu.cn}
\affiliation{International Center for Quantum Materials, School of Physics, Peking University, Beijing 100871, China}
\affiliation{Beijing Academy of Quantum Information Sciences, Beijing 100913, China}

\begin{abstract}
Searching for Kitaev quantum spin liquid (QSL) is a fascinating and challenging problem. Much effort has been devoted to honeycomb lattice candidates with strong spin-orbit coupling in 5d-electron iridates and 4d-electron RuCl$_{3}$. Recently, theoretical studies suggested that the $3d^{7}$ Co-based honeycomb materials with high spin state S=3/2 and effective orbital angular momentum L=1 could also be promising candidates of Kitaev QSL. One of the candidates, BaCo$_{2}$(AsO$_{4}$)$_{2}$, was revisited recently. The long range magnetic order in BaCo$_{2}$(AsO$_{4}$)$_{2}$ can be suppressed by very weak in-plane magnetic field, suggesting its proximity to Kitaev QSL. Here we perform time domain terahertz spectroscopy measurement to study the magnetic excitations on BaCo$_{2}$(AsO$_{4}$)$_{2}$. We observe different magnon excitations upon increasing external magnetic field. In particular, the system is easily driven to a field-polarized paramagnetic phase, after the long range magnetic order is suppressed by a weak field $H_{c2}$. The spectra beyond $H_{c2}$ are dominated by single magnon and two magnon excitations without showing signature of QSL. We discuss the similarity and difference of the excitation spectra between BaCo$_{2}$(AsO$_{4}$)$_{2}$ and the widely studied Kitaev QSL candidate RuCl$_{3}$.

\end{abstract}

\maketitle
\section{Introduction}
Quantum spin liquids (QSLs) are novel states of magnetic systems characterized by the absence of long range spin order down to zero temperature\cite{Balents2010,RevModPhys.89.025003,Savary_2016}. Amongst various QSLs, the Kitaev QSL based on honeycomb lattice is of especial importance. Different from spin liquids arising from geometrically frustrated spin arrangements, the bond-dependent Kitaev spin interactions frustrate the spin configuration. The Kitaev model hosts exactly solvable QSL ground state and fractionalized excitations described by Majorana fermions\cite{KITAEV20062}. Experimentally, the Kitaev model is expected to be realized in honeycomb Mott insulators with spin-orbit coupling\cite{PhysRevLett.102.017205}. Much effort has been made to experimentally explore materials dominated by the bond dependent Kitaev interaction, firstly in transition metal 5d-electron iridates and then in 4d-electron RuCl$_{3}$. However, there are significant non-Kitaev interactions in those real materials, such as the Heisenberg type exchange interaction and off-diagonal exchange interactions. The non-Kitaev interactons hinder the formation of pure Kitaev quantum spin liquid and push the system to be ordered at low temperature\cite{PhysRevLett.112.077204,PhysRevB.94.064435}. To approach the Kitaev QSL, one promising route is to suppress the long range magnetic order by applying magnetic field. The field induced quantum spin liquid candidate states are widely studied in these Kitaev materials\cite{Ruiz2017,PhysRevLett.110.097204,PhysRevLett.117.277202,Banerjee2016,Banerjee1055}.

Recently, the $3d^{7}$ Co-based honeycomb materials with high spin state S=3/2 and effective orbital angular momentum L=1 have been proposed as new candidates of Kitaev QSLs and have attracted wide attention \cite{PhysRevB.97.014407,PhysRevB.97.014408,PhysRevLett.125.047201,PhysRevB.97.134409,REGNAULT1977660,10.1016/j.heliyon.2018.e00507,REGNAULT1979194,VICIU20071060,PhysRevMaterials.3.074405,PhysRevB.102.224429,PhysRevB.101.085120,PhysRevB.102.054414,PhysRevB.102.224411,PhysRevB.94.214416,PhysRevB.103.L180404,doi:10.1063/1.5029090,2012.00940v2,Motome_2020,2106.11982v1}. In transition metal 4d- and 5d-electron systems, the spatially extended d-electron wave function leads to nonnegligible longer range coupling, which is detrimental to Kitaev QSL. Comparing with 4d- and 5d-electron systems, the 3d-electrons have more localized wave functions, thus recede the longer range coupling. Besides, the Heisenberg interactions in 3d systems are easier to be minimized by tuning the external parameters. Accordingly, the $3d^{7}$ systems may be more appropriate to realize the Kitaev QSL. BaCo$_{2}$(AsO$_{4}$)$_{2}$ is among the suggested candidates with $3d^{7}$ Co honeycomb lattice. It has similar properties as the well-established Kitaev QSL candidate RuCl$_{3}$. The magnetic susceptibility shows strong anisotropy, indicating the anisotropic exchange interactions in the spin Hamiltonian. Especially, a small magnetic field applied in the honeycomb plane can significantly change the magnetic system and completely suppress the long range order. The critical field $\sim$0.5 T is much weaker compare to that of 7 T in RuCl$_{3}$, indicating very small Heisenberg interaction\cite{Zhongeaay6953}. Those studies suggest that this system is an excellent candidate for realization of a field-induced QSL state. To characterize the magnetic excitations at low energies, we perform terahertz (THz) time domain spectroscopy measurement on BaCo$_{2}$(AsO$_{4}$)$_{2}$ crystals under external magnetic field. We monitored the evolution of the spin wave under magnetic field and characterized the emergent magnetic excitations in the field induced paramagnetic state.

\section{Sample and experiments}

The single crystals of BaCo$_{2}$(AsO$_{4}$)$_{2}$ with a typical size of $3\times3\times0.3 mm^{3} $ were grown by the flux method \cite{Zhongeaay6953}. The compound crystallizes in the trigonal centrosymmetric space group R-3. The honeycomb structure is made of edge sharing CoO$_6$ tetrahedra and stacked along the c-axis with an ABC periodicity. Below 5.4 K, the system transforms to an antiferromagnetic ordered state. The magnetic order is rather complex with spiral spin chains \cite{10.1007/978-94-009-1860-3}. By applying in-plane magnetic field, the system exhibit two phase transitions near 0.2 T and 0.5 T, respectively. The antiferromagnetic long range order is suppressed at the second transition, while the first transition is more complicated \cite{Zhongeaay6953}.

Time domain THz transmission spectra were measured by using a home built spectroscopy system equipped with helium cryostat and Oxford spectramagnet \cite{PhysRevB.98.094414}. As shown in Figure~\ref{Fig:1} (a), the wave vector of the incident THz beam is perpendicular to the crystallographic ab-plane. The polarization of the magnetic field component of THz wave can be tuned from a-axis to b-axis. By rotating the superconducting magnet, the external magnetic field can be applied either parallel or perpendicular to the ab-plane. The time domain signals of the sample and reference (empty aperture) were detected via free space electro-optics sampling in ZnTe crystal. Fourier transformation of the time domain spectra provides the frequency dependent complex transmission spectra containing both magnitude and phase information, from which the real and imaginary parts of optical constants can be extracted (see Supplementary Material)\cite{sup}.

\section{Evolution of the magnon mode with temperature and magnetic field in the antiferromagnetic order region}

\begin{figure}[htbp]
	\centering
	\includegraphics[width=8cm]{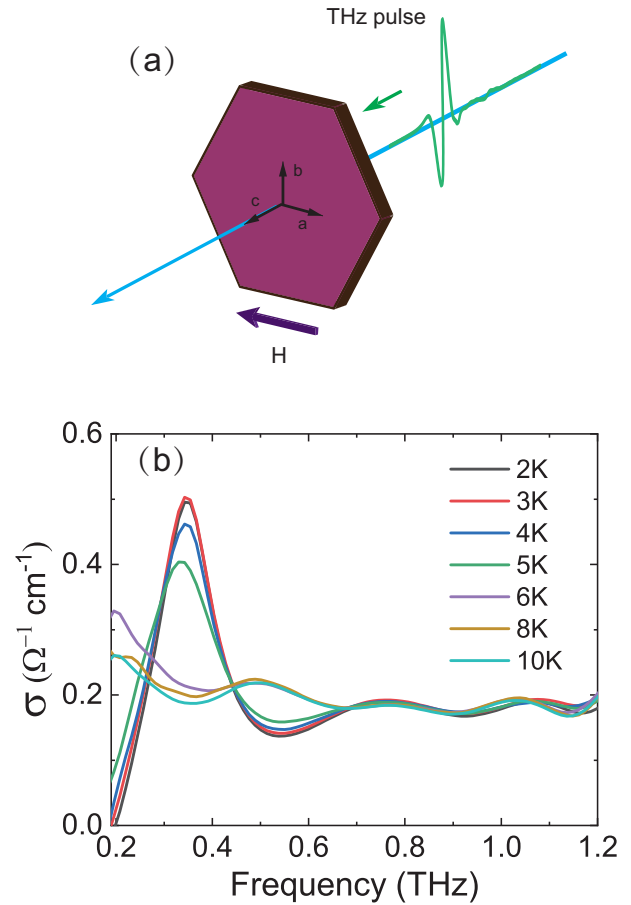}\\
	\caption{(a) Schematic of the terahertz transmission spectra measurement.   (b) Real part of the optical conductivity at selected temperatures without external field.
	}\label{Fig:1}
\end{figure}

Figure~\ref{Fig:1} (b) shows the temperature dependence of the real part of the optical conductivity $\sigma$ at low temperatures. Below $T_N$, a narrow peak emerges abruptly on the flat background at 0.35 THz. Then, its intensity gradually increases and reaches a maximum at $\sim$3 K. We label this peak as mode A. The frequency of the mode is consistent with previous neutron scattering experiments, being identified as the magnon at the $\Gamma$ point of the magnetically ordered state \cite{10.1007/978-94-009-1860-3,10.1016/j.heliyon.2018.e00507}. In time domain THz measurement, ordered spins are excited by THz wave through Zeeman torque $dM/dt = \gamma M \times H_{THz}$, where $\gamma$ denotes the gyromagnetic constant, $M$ the magnetic momentum and $H_{THz}$ the magnetic field component of the THz wave. To effectively drive the spin procession, \emph{i.e.} excite magnon, $H_{THz}$ should be perpendicular to the magnetic moment. Therefore, measuring the polarization dependence of the magnon can help identify the spin orientation. When we rotate the polarization of magnetic field component of the THz wave from a- to b-axis without applying external magnetic field, the magnon mode is visible in all directions. The result is consistent with the antiferromagnetic order with presence of spiral spin chains \cite{10.1007/978-94-009-1860-3}.

\begin{figure}[htbp]
	\centering
	\includegraphics[width=9cm]{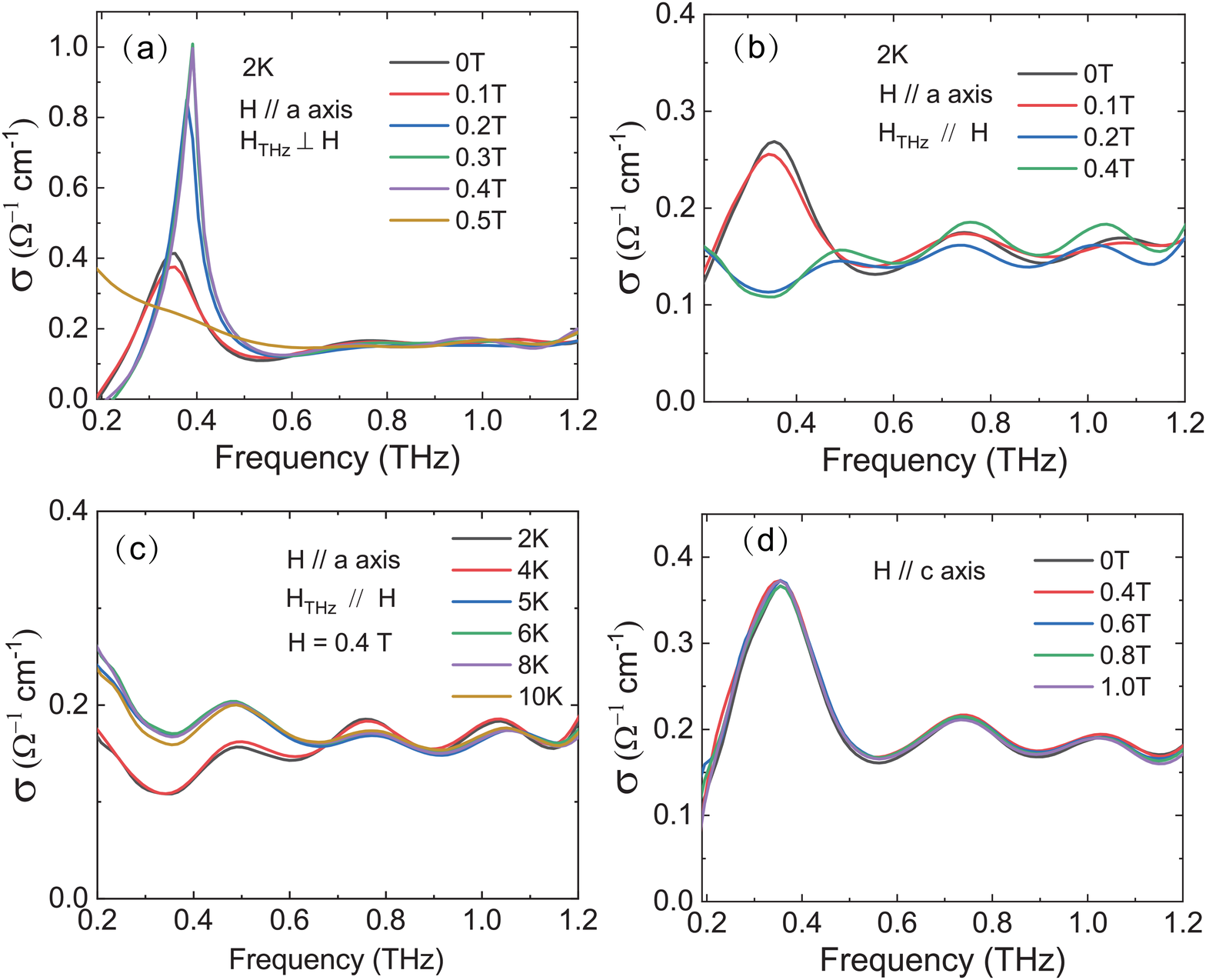}\\
	\caption{ Evolution of the optical conductivity under magnetic field at 2 K. (a) Spectra for H along a axis and $H_{THz}$ along b axis.  (b) Spectra for H along a axis and $H_{THz}$ along a axis. (c) Temperature dependence of the spectra for H (0.4 T) along a axis and $H_{THz} \parallel H$. (d) Spectra for H along c axis.
	 }\label{Fig:2}
\end{figure}

We now present the evolution of THz spectra under applied external magnetic field. Figure~\ref{Fig:2} (a) shows the frequency-dependent conductivity measured at 2 K in the configuration of the external magnetic field $H$ being applied along the a-axis and the magnetic field of THz wave perpendicular to $H$. Upon increasing the magnetic field, the intensity of the magnon mode A initially decreases slightly, but keeps the mode frequency unchanged. At around $H_{c1}$ = 0.2 T, the excitation mode suddenly shifts to 0.39 THz, and its intensity is strongly enhanced, suggesting a magnetic field induced phase transition above 0.2 T. We denote this sharp peak as mode B for the field above 0.2 T. As the magnetic field increases to 0.5 T, the narrow peak excitations are fully suppressed. The results are consistent with previous reports that a magnetic field induced transition from antiferromagnetic phase I to phase II occurs near 0.2 T and a complete suppression of antiferromagnetic order \cite{10.1007/978-94-009-1860-3,Zhongeaay6953}.

The spectra with $H_{THz}\parallel H$ are quite different. As shown in Fig.~\ref{Fig:2} (b), in the antiferromagnetic phase II above 0.2 T, the excitation mode at 0.39 THz is completely absent. Instead, a slight enhancement in conductivity spectra above 0.7-0.8 THz seems to be visible. Figure~\ref{Fig:2} (c) shows the temperature dependent spectra measured at 0.4 T in the antiferromagnetic phase II region. We do not observe any excitation peak emerging below T$_N$. The low energy spectral weight drops below T$_N$ and reduced spectral weight is shifted to the higher energy above 0.7-0.8 THz. Based on those results, we identify that the first phase transition near 0.2 T is a spin reorientation transition. The moments in spiral spin chains are tuned by the magnetic field. The antiferromagnetic phase II is consistent with a collinear antiferromagnetic state, consistent with previous report\cite{Zhongeaay6953}. The disappearance of magnon mode in Fig.~\ref{Fig:2} (c) can be attributed to the polarization selection rule of $H_{THz}$ parallel to the magnetic moments. For a comparison, the magnetic state is stable when the external magnetic field is applied along the c-axis, as shown in Fig.~\ref{Fig:2} (d). Those results are consistent with static measurement \cite{10.1007/978-94-009-1860-3,Zhongeaay6953}.

\begin{figure}[htbp]
	\centering
	\includegraphics[width=9cm]{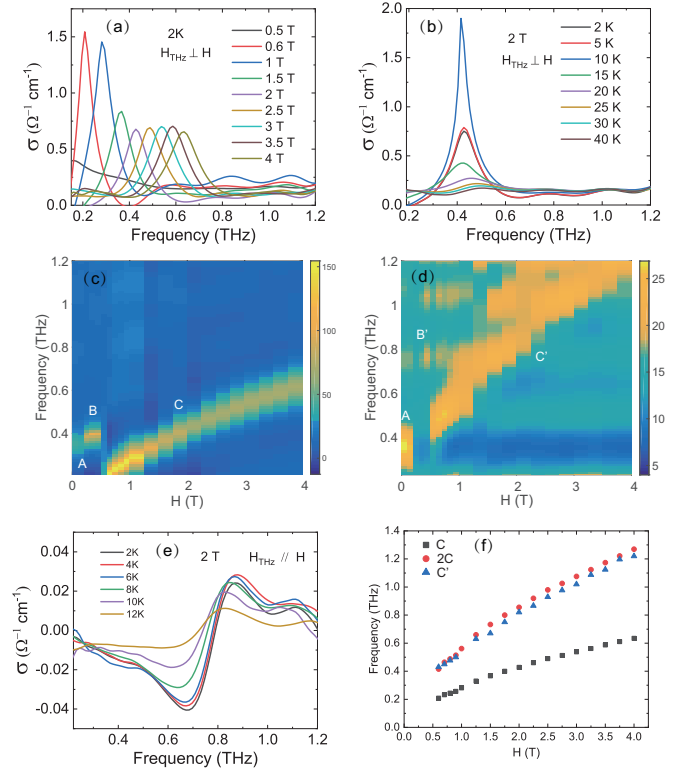}\\
	\caption{  (a) Real part of the optical conductivity at different magnetic field for H along a axis and $H_{THz}$ along b axis.  (b) Temperature evolution of the excitation at 2 T for $H_{THz} \perp H$. (c)(d) Contour plot of $\sigma$ as a function of field and frequency for $H_{THz} \perp H$ and $H_{THz} \parallel H$. (e) Temperature dependence of the spectra at 2 T for $H_{THz} \parallel H$. The spectrum at 15 K are subtracted as a reference. (f) Field dependence of the excitation modes.	}\label{Fig:3}
\end{figure}

\section{Magnetic excitations in the field induced paramagnetic state}

Across the second critical field $H_{c2}$ at around 0.5 T, the sample transforms to a paramagnetic phase. Identifying the nature of the excitations in this phase is essential to judge whether or not it is a quantum spin liquid state. The measured spectra of this phase are shown in Fig.~\ref{Fig:3}. The field evolution of the spectra with
$H_{THz}\perp H$ are shown in Fig.~\ref{Fig:3} (a). All the well defined modes are fully suppressed at 0.5 T. Upon further increasing the in-plane magnetic field slightly, a new narrow peak emerged immediately at $\sim$0.6 T, denoted as mode C. This peak has the largest intensity in our measurement. Its frequency increases with the magnetic field. Obviously, mode C is the magnon in field-polarized paramagnetic phase, being essentially the same as the ferromagnetic order. Surprisingly, we find an anomalous behavior for this mode as shown in Fig.~\ref{Fig:3} (a). At 2 K, the intensity of the mode drops dramatically when the magnetic field is above 1 T. To track the characteristics of the mode, we measured the temperature dependence of the spectra at each field. The typical behavior of the mode at relatively high field, \emph{e.g.} 2 T, are shown in Fig.~\ref{Fig:3} (b). The temperature evolution of the mode intensity shows an unusual strong resonance character. Below 1 T, the resonance temperature is around 2 K, the mode decreases monotonically as temperature increases. With increasing magnetic field, the resonance temperature also increases. At 2 T, the mode intensity gets maximum at around 10 K, while the intensity at 2 K becomes much smaller. It is reasonable to attribute the resonance behavior of the mode to the competing interactions among the exchange interaction, magnetic field and thermal excitations.

Figure~\ref{Fig:3} (c) displays the intensity plot of the mode frequency as a function of external magnetic field in the configuration of $H_{THz}\perp H$ at 2 K, in which we can identify clearly three magnon modes upon increasing the magnetic field. Mode A and B are magnons in antiferromagentic phase I and II, respectively, while mode C is the magnon in field-polarized paramagnetic phase.

Similarly, we plot the intensity map of the conductivity spectra for $H_{THz} \parallel H$ in Fig.~\ref{Fig:3} (d). Note that the intensity of the excitation spectrum is much weaker than that observed for $H_{THz} \perp H$. In order to clearly identify the magnetic excitations, we substrate the spectrum at 15 K as a reference. The typical spectra at a representative field, \emph{e.g.} 2 T, is shown in Fig.~\ref{Fig:3} (e). We emphasize that the mode C observed with $H_{THz} \parallel H$ is completely absent here, indicating that this mode also follows the polarization selection rules. On the contrary, another broad mode feature at higher energy, labelled as mode C', is observed with a dip at lower energy side and a broad peak at higher energy side. The temperature evolution is monotonous for mode C' in all applied magnetic field above 0.6 T. The characteristic behaviors imply an opening of energy gap below the mode C' for $H_{THz} \parallel H$.

To further identify the relation of magnetic excitations between $H_{THz} \perp H$ and $H_{THz} \parallel H$, we plot the peak positions of mode C and C' in Fig.~\ref{Fig:3} (f). We noticed that both have a sublinear field dependence, and mode C' has approximately the energy twice of that of mode C. The doubled frequency of mode C is also plotted as the circles in Fig.~\ref{Fig:3} (f), showing a good match with mode C'. In field-polarized paramagnetic phase, the magnetic moment in each spin chain is along the external field direction. With $H_{THz} \perp H$, the THz wave coherently drives the spin procession and excites the single magnon. While for $H_{THz} \parallel H$, the THz wave can not excite the single magnon, but excite the two magnons which show an energy gap below the mode and a continuum structure, as we shall explain below.

\begin{figure}[htbp]
	\centering
	\includegraphics[width=7.5cm]{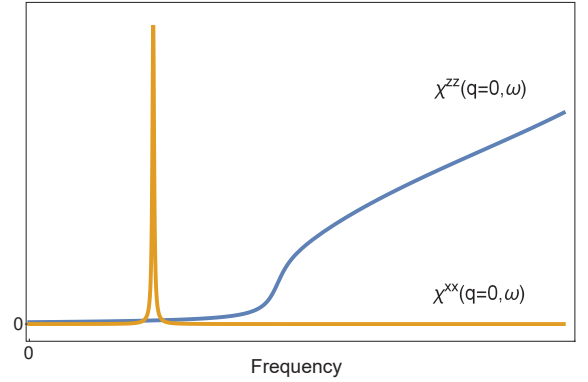}\\
	\caption{The calculated dynamical spin structure factors $\chi^{xx}({\boldsymbol{q}}=0,\omega)$ (orange line, for $H_{THz}\perp H$ case) showing a single magnon peak, and $\chi^{zz}({\boldsymbol{q}}=0,\omega)$ (blue line, for $H_{THz}\parallel H$ case) showing two-magnon continuum whose low energy edge is twice of the single magnon energy. Details about this calculation can be found in Supplementary Materials \cite{sup}.}\label{Fig:4}
\end{figure}

To understand the physics begetting the novel spin excitation spectrum of BaCo$_{2}$(AsO$_{4}$)$_{2}$, we consider a simplified model on honeycomb lattice and show that its excitation spectrum qualitatively reproduces the observed mode C and C' in the high field phase.
The model Hamiltonian is
\beq
\hat{H}=-J\sum_{<ij>} \boldsymbol{S}_i\cdot \boldsymbol{S}_j-g_z\mu_B H\sum_{i}S_i^{z},
\eeq
where $J$ is the nearest-neighbor ferromagnetic Heisenberg exchange coupling, $g_z$ is the gyromagnetic ratio, $\mu_B$ is the Bohr magneton, $H$ is the applied external magnetic field. Without loss of generality we define the $z$ direction along the applied field $H$, and the $x$ direction along the $H_{THz}$ direction of THz wave that is perpendicular to applied field $H$. The measured THz spectra is then essentially the dynamical spin structure factors $\chi^{xx}(\boldsymbol{q},\omega)$ and $\chi^{zz}(\boldsymbol{q},\omega)$ for the $H_{THz}\perp H$ and $H_{THz}\parallel H$ cases respectively.
The wave vector ${\boldsymbol{q}}$ of the THz wave is much smaller than the Brillouin zone size and is treated as ${\boldsymbol{q}}=0$ hereafter.
We compute the magnon dispersion and dynamical spin structure factors for this model by the standard linear spin wave theory, the calculation details can be found in Supplementary Materials \cite{sup}.
This model has a polarized ground state and two branches (because of the two-site unit cell of honeycomb lattice) of gapped magnons with dispersion
\beq
\varepsilon_{\boldsymbol{q}}^{\pm}=g_z\mu_B H+\frac{3J}{2}\left ( 1\pm \left\vert \frac{1}{3}\sum_{\delta_{1,2,3}}e^{i {\boldsymbol{q}}\cdot {\boldsymbol{\delta}}} \right\vert \right ),
\eeq
where $\boldsymbol{\delta}_{1,2,3}$ are the three bond vectors connecting nearest-neighbor sites on the honeycomb lattice.
The magnon dispersions $\varepsilon_{\boldsymbol{q}}^{\pm}$ depend on external field $H$ linearly.
The calculated $\chi^{xx}({\boldsymbol{q}}=0,\omega)$ is
\beq
\chi^{xx}({\boldsymbol{q}}=0,\omega)\propto \delta(\omega -\varepsilon_{\boldsymbol{q}}^{-}),
\eeq
and shows a sharp peak at single magnon energy
$\varepsilon_{\boldsymbol{q}=0}^{-}=g_z\mu_B H$.
The calculated $\chi^{zz}({\boldsymbol{q}}=0,\omega)$  is
\beq
\chi^{zz}({\boldsymbol{q}}=0,\omega)\propto \sum_k [\delta(\omega -\varepsilon_{\boldsymbol{k}}^{+}-\varepsilon_{-{\boldsymbol{k}}}^{+})+\delta(\omega -\varepsilon_{\boldsymbol{k}}^{-}-\varepsilon_{-{\boldsymbol{k}}}^{-})],
\eeq
and shows broad two-magnon continuum.
Fig.~\ref{Fig:4} shows an example of the calculated dynamical spin structure factors.
The low energy edge of the continuum in $\chi^{zz}({\boldsymbol{q}}=0,\omega)$ is twice of the minimal single magnon energy $\omega=2\varepsilon_{{\boldsymbol{k}}=0}^-=2g_z\mu_B H$.
The spin model for BaCo$_{2}$(AsO$_{4}$)$_{2}$ will certainly be much more complicated than the Heisenberg model considered here\cite{10.1007/978-94-009-1860-3}. But its dynamical spin structure factors in the high field polarized phase will have the same qualitative behaviors, namely that $\chi^{xx}$ shows a single magnon peak
and $\chi^{zz}$ shows a two magnon continuum whose low energy edge is twice of the minimal single magnon energy.

Based on above discussions, we summarize the assignment of magnetic excitations displayed in Fig. \ref{Fig:3} (c) and (d). The mode A at 0.35 THz below 0.2 T is the magnon excitation in antiferromagnetic phase I. It is observed for both $H_{THz} \perp H$ and $H_{THz} \parallel H$ due to the presence of spiral spin chains. The mode B at 0.39 THz above 0.2 T but below 0.5 T is the magnon excitation in antiferromagnetic phase II. It is observed only for $H_{THz} \perp H$. There is a very weak feature near 0.8 THz for $H_{THz} \parallel H$, denoted as B', which is likely originated from the two-magnon excitations in the field-induced collinear antiferromagnetic phase II. The mode C and C' observed for $H_{THz} \perp H$ and $H_{THz} \parallel H$ respectively are single magnon and two-magnon excitations in field-polarized paramagnetic phase. The reason that the single magnon excitation is absent for $H_{THz} \parallel H$ in antiferromagnetic phase II and field-polarized paramagnetic state is because the magnetic component of THz wave is parallel to the moment orientation in both phases.

\section{Further discussions on the field induced paramagnetic state}

Our study indicates that BaCo$_{2}$(AsO$_{4}$)$_{2}$ offers an ideal system to investigate the magnetic excitations in
$3d^{7}$ Co-based honeycomb lattice systems. The external magnetic field required to suppress the antiferromagnetic order is much smaller than that in any other known magnetic honeycomb compounds, enabling a full and careful characterization of magnetic excitations by time domain THz spectroscopy technique.

It is also interesting to note that the present measurement result on BaCo$_{2}$(AsO$_{4}$)$_{2}$ shares similarity to the well-studied RuCl$_3$ compound under high magnetic field. As mentioned above, both the mode C and mode C' in the field induced paramagnetic state have a sublinear field dependence, with the slopes gradually decreasing with the field increasing. These modes seem to have the similar characters with the excitations observed in the field induced disorder state of RuCl$_3$ \cite{PhysRevLett.119.227202,PhysRevB.96.241107,PhysRevLett.125.037202,PhysRevB.101.140410,Wulferding2020}.

Although the compound BaCo$_{2}$(AsO$_{4}$)$_{2}$ was suggested to be a possible new Kitaev QSL candidate when the magnetic order was suppressed by the magnetic field of 0.5 T \cite{Zhongeaay6953}, our study revealed sharp single magnon and two magnon excitations even the in-plane magnetic field is as low as 0.6 T, which suggests against formation of Kitaev QSL state. The sharp magnon excitation is absent only in a region when the magnetic field is extremely close to 0.5 T.

\section{Conclusion}
In conclusion, we have presented THz spectroscopy study on Co-based honeycomb BaCo$_{2}$(AsO$_{4}$)$_{2}$ under in-plane magnetic field up to 4 T. The fact that an extremely small magnetic field can suppress the long range antiferromagnetic phase makes it an ideal system to investigate the magnetic excitations in the field induced states by THz spectroscopy measurement. Our field and polarization dependent measurement reveals two first order transitions. The first transition at 0.2 T is from spiral order to collinear antiferromagnetic order. The second transition at 0.5 T is the suppression of the antiferromagnetic order. We observed different magnon excitations in different regions of applied magnetic field. In particular, after the long range magnetic order was suppressed by a weak field $H_{c2}$, the system was driven immediately to a field-polarized paramagnetic phase being similar to a ferromagnetic state. The spectra beyond $H_{c2}$ are dominated by single magnon and two magnon excitations. However, no signature of a quantum spin liquid state is observed. We also compared the excitation spectra of BaCo$_{2}$(AsO$_{4}$)$_{2}$ with that of widely studied 4d-electron Kitaev candidate RuCl$_{3}$ and addressed their similarities and differences in magnetic excitations.

\begin{center}
\small{\textbf{ACKNOWLEDGMENTS}}
\end{center}

This work was supported by National Natural Science Foundation of China (No. 11888101), the National Key Research and Development Program of China (No. 2017YFA0302904). The crystals were grown in the laboratory of R.J. Cava at Princeton University.

\bibliographystyle{apsrev4-1}

\bibliography{BCAO_database}

\end{document}